# ITRUSST Consensus on Biophysical Safety for Transcranial Ultrasound Stimulation

Jean-François Aubry [1], David Attali [1,2], Mark E. Schafer [3], Elsa Fouragnan [4], Charles F. Caskey [5], Robert Chen [6,7], Ghazaleh Darmani [7], Ellen J. Bubrick [8], Jérôme Sallet [9,10], Christopher R. Butler [11,12], Charlotte J. Stagg [13,14], Miriam C. Klein-Flügge [10,15], Seung-Schik Yoo [16], Christy Holland [17,18], Brad Treeby [19,20], Eleanor Martin [19], Lennart Verhagen [21]*, Kim Butts Pauly [22]*
* These authors contributed equally

[1] Physics for Medicine Paris, Inserm U1273, ESPCI Paris, CNRS UMR8063, PSL University, Paris, France

[2] GHU-Paris Psychiatrie et Neurosciences, Hôpital Sainte Anne, Université Paris Cité, 75014 Paris, France

[3] Rockefeller Neuroscience Institute, West Virginia University, Morgantown, West Virginia, USA

[4] Brain Research Imaging Center, 7 Derriford Rd, Plymouth PL6 8BU; School of Psychology, Portland Square, Plymouth PL4 8AA, UK

[5] Vanderbilt University Institute of Imaging Sciences, VU Medical Center, Nashville, Tennessee, USA

[6] Division of Neurology, Department of Medicine, University of Toronto, Toronto, Ontario, Canada

[7] Krembil Research Institute, University Health Network, Toronto, Ontario, Canada

[8] Brigham and Women's Hospital, Harvard Medical School, Department of Neurology, 75 Francis St., Boston, MA, USA

[9] Univ Lyon, Université Lyon 1, Inserm, Stem Cell and Brain Research Institute U1208, Bron, France

[10] Oxford Centre for Integrative Neuroimaging (OxCIN), Department of Experimental Psychology, University of Oxford, Oxford, OX1 3TA, UK

[11] Department of Brain Sciences, Imperial College London, London, W12 0BZ, UK

[12] The George Institute for Global Health, Imperial College London, London, W12 7RZ, UK

[13] Oxford Centre for Integrative Neuroimaging (OxCIN), FMRIB, Nuffield Department of Clinical Neurosciences, University of Oxford, FMRIB Building, John Radcliffe Hospital, Headington, Oxford, OX3 9DU, UK

[14] Medical Research Council Brain Network Dynamics Unit, Nuffield Department of Clinical Neurosciences, University of Oxford, Mansfield Road, Oxford, OX1 3THU, UK

[15] Department of Psychiatry, Warneford Hospital, University of Oxford, Oxford OX3 7JX, UK

[16] Brigham and Women's Hospital, Harvard Medical School, Department of Radiology, 75 Francis St., Boston, MA, USA

[17] Division of Cardiovascular Health and Disease, Department of Internal Medicine, University of Cincinnati, Cincinnati, Ohio, USA
[18] Department of Biomedical Engineering, University of Cincinnati, Cincinnati, Ohio, USA
[19] Department of Medical Physics and Biomedical Engineering, University College London, London, UK

[20] NeuroHarmonics LTD, 71-75 Shelton Street, Covent Garden, London, UK

[21] Donders Institute for Brain, Cognition and Behaviour, Radboud University, 6525 GD Nijmegen, the Netherlands

[22] Stanford University, Department of Radiology, Stanford CA 94305, USA

*email: kimbutts@stanford.edu; jean-francois.aubry@espci.fr; lennart.verhagen@donders.ru.nl*

## Abstract

Transcranial ultrasound stimulation (TUS) is an emerging technology for non-invasive brain stimulation. The International Transcranial Ultrasonic Stimulation Safety and Standards consortium (ITRUSST) has established consensus on considerations for nonsignificant biophysical risk of TUS, drawing upon the literature and established regulations for biomedical devices. Here, we assume the application of TUS to individuals without contraindications, compromised thermoregulation, vascular vulnerabilities, or administered ultrasound contrast agents. In this context, we present a concise yet comprehensive set of levels for nonsignificant risks of TUS application. For mechanical effects, it is non-significant risk if the mechanical index (MI) or the mechanical index for transcranial application ($MI_{tc}$) does not exceed 1.9. For thermal effects, it is non-significant risk if any of the following three levels are met: the peak temperature rise does not exceed 2°C or the peak absolute temperature does not exceed 39°C, assuming a baseline temperature of 37°C, the thermal dose does not exceed 2 CEM43 in brain tissue, 16 CEM43 in bone tissue, and 21 CEM43 in skin tissue, or specific values of the thermal index (TI) for a given exposure time. This report reflects a consensus expert opinion and can inform, but not replace, regulatory guidelines or official international standards. Similarly, this consensus can inform, but not replace, ethical evaluation, which weighs the total burden, risks, and benefits of the proposed action. The stated levels are not safety limits per se, and further data is needed to establish the threshold for significant risk. We review literature relevant to our considerations and discuss limitations and future developments of our approach.

## Highlights

- Mechanical risks of TUS are nonsignificant if either MI or MItc does not exceed 1.9.
- Thermal risks are nonsignificant if the absolute temperature does not exceed 39 °C, assuming a baseline temperature of 37°C.
- Thermal risks are nonsignificant if the thermal dose does not exceed 2 CEM43 in brain.
- Thermal risks are nonsignificant for specific thermal index exposure durations.
- The ITRUSST consensus is that these levels present non-significant risk. Protocols exceeding these levels are not necessarily unsafe or significant risk. Further data is needed to establish the threshold for significant risk.

# 1. Introduction

Transcranial ultrasound stimulation (TUS) is an emerging technology for non-invasive brain stimulation. Currently, there are no established guidelines for the safe application of ultrasound neuromodulation in humans. Therefore, in a series of meetings and benefitting from feedback on initial drafts, the International Transcranial Ultrasonic Stimulation Safety and Standards consortium (ITRUSST) has established consensus on considerations for the biophysical safety of TUS. The current report is intended to help investigators, operators, and manufacturers adjust ultrasound parameters within non-significant risk biophysical levels, ensuring minimal and nonsignificant risk for structural damage. This report reflects a consensus expert opinion and can inform but not replace regulatory guidelines or official international standards. Similarly, the current report can inform but does not replace the risk/benefit and ethical evaluation by institutional review boards. These boards generally consider multiple aspects of the proposed action, also, for example, precision of the targeting, inclusion/exclusion criteria, burden of participation, potential interactions, physiological effects, and informed consent. These important considerations fall outside the scope of the current report; here, we focus on the biophysical safety of TUS.

There are two main biophysical risks associated with the application of ultrasound: mechanical and thermal bioeffects [1]. Mechanical bioeffects mainly concern the risk of acoustic cavitation, which can lead to local tissue damage such as cell death or blood vessel hemorrhage. Thermal bioeffects may occur when mechanical energy is converted into thermal energy through absorption, leading to tissue heating and potential thermal damage.

The pulses discussed in this document for TUS are similar to those used in diagnostic ultrasound in terms of pressure amplitude but differ significantly in frequency and pulse duration. TUS is typically in the range of 200-800 kHz, while for diagnostic ultrasound, fundamental frequencies are often in the range of 1-10 MHz. The pulse durations, which are typically longer for TUS, at 100 microseconds to 100 ms or longer, compared to those in diagnostic ultrasound, which could be 3 to 160 microseconds. These differences inform the mechanical and thermal considerations for the biophysical safety of TUS and could also impact which existing regulatory guidelines are most relevant for this application.

ITRUSST brings together researchers, manufacturers, regulators, funders, and other experts and stakeholders to advance the safe and effective application of transcranial focused ultrasound for neuromodulation. ITRUSST is endorsed by the Focused Ultrasound Foundation (FUSF) and the International Federation of Clinical Neurophysiology (IFCN) to provide best practices, guidelines, and consensus on TUS [2,3]. Here, we draw upon existing regulatory guidelines and present a concise yet comprehensive set of parameters and levels that we consider biophysically non significant risk, as summarized in Table 1. Notably, we do not define levels above which TUS has a high probability of leading to structural damage (i.e., a safety limit); instead, we attempt to define levels below which there is abundant evidence that no significant biophysical harm is likely to occur. Application beyond these levels and assumptions does not imply a significant risk per se, but, at this point in time, there is insufficient data to make a specific recommendation at a level above MI=1.9 that would establish a non-significant risk threshold.

The ITRUSST safety group reviewed the literature and considered the relevance of established standards and guidelines for biomedical devices, including diagnostic ultrasound, from organizations such as the International Organization for Standardization (ISO), the International Electrotechnical Commission (IEC), the Food and Drug Administration (FDA), the British Medical Ultrasound Society (BMUS), and the American Institute of Ultrasound in Medicine (AIUM). The resultant consensus solely addresses low-intensity transcranial ultrasound for neuromodulation (TUS). The current consensus explicitly does not consider transcranial high-intensity focused ultrasound (HIFU) for hyperthermia and ablation. To give an example of the order of magnitude difference between HIFU and TUS, HIFU for ablation might use 500 W/cm$^2$ for 20 seconds (10,000 Joules) [4], while one example set of TUS pulses used 20 W/cm$^2$ for 80 s at a duty cycle of 10% (160 Joules) [5]. Additionally, thermal HIFU treatments usually induce temperatures of 55°C or more [6], sustained for a few seconds to cause cell death, and thermal lesioning temperatures reported in ultrasound thalamotomy are in the range 55-60 °C [7], which contrast with the thermal limits suggested here. The current consensus also explicitly does not consider focused ultrasound with administered microbubbles for blood-brain barrier opening (BBBO) [8], or non-transcranial applications, such as peripheral and retinal neuromodulation. We restrict our statement on non-significant risk to the frequency range in which TUS is typically applied, between 200 and 800 kHz. We assume the application of TUS to persons with uncompromised thermoregulation, without vascular vulnerabilities, without contraindications, and in the absence of cavitation nuclei, such as ultrasound contrast agents.

To avoid redundancy in definitions, we reference a recent publication on the standardized reporting of TUS [2], which includes all definitions of variables used here.

*Table 1: Summary of the parameters and levels ITRUSST considers non-significant risk. \*One should determine which mechanical and thermal indices (TI) correspond best to the applied configuration (MI or MI$_{tc}$, see section 2.1; and TIS, TIB, or TIC for the measure of TI, see section 2.2.2). For many TUS applications, the MI$_{tc}$ and the TIC will be most appropriate.*

| Bioeffects | ITRUSST nonsignificant risk consensus |
|---|---|
| **Mechanical** | Any of:*<br>• MI ≤ 1.9<br>• MI$_{tc}$ ≤ 1.9 |
| **Thermal** | Any of:<br>• Temperature rise ≤ 2°C<br>• Absolute temperature ≤ 39° Celsius, assuming a baseline temperature of 37°C<br>• Thermal dose depending on the tissue type<br>    ≤ 2 CEM43 for brain tissue<br>    ≤ 16 CEM43 for bone tissue<br>    ≤ 21 CEM43 for skin tissue<br>• Maximum exposure time depending on the level of TI*:<br>    80 min   1.5 < TI ≤ 2.0<br>    40 min   2.0 < TI ≤ 2.5<br>    10 min   2.5 < TI ≤ 3.0<br>    160 sec  3.0 < TI ≤ 4.0<br>    40 sec   4.0 < TI ≤ 5.0 |

|  | 10 sec | 5.0 < TI ≤ 6.0 |

## 2. Current relevant regulatory guidelines
### 2.1. Mechanical safety

To estimate the risk of mechanical bioeffects from any diagnostic ultrasound protocol, the Mechanical Index (MI) was adopted by the FDA. The MI was designed to inform clinical users about potential mechanical effects during diagnostic ultrasound and was derived from an analysis of the minimum rarefactional pressure necessary to cause inertial cavitation, in the presence of cavitation nuclei in water. MI is defined by the spatial-peak value of the peak-rarefactional pressure measured in water and given in MPa, derated by 0.3 dB/cm/MHz, divided by the square root of the center frequency given in MHz [2,11,12]. MI can be calculated from device settings alone, without further knowledge of the application.

The derating of 0.3 dB/cm/MHz in the definition of MI was established to be conservative for many body applications, but it does not consider the insertion loss presented by the skull bone. If the MI is estimated using this standard derating factor, as per the regulatory parameters for diagnostic ultrasound, the pressure inside the brain is overestimated. The diagnostic MI guidelines are, therefore, overly conservative for transcranial applications of ultrasound. Alternatively, derating can be accounted for more precisely using a transcranial application-specific MI, which we propose to call $MI_{tc}$ (mechanical index for transcranial application), described in more detail in section 3.1 and the ITRUSST standardized reporting guidelines [2].

The FDA regulatory limit is an MI ≤ 1.9 for all applications except the eye [11]. Experimental work suggests that significantly higher rarefactional pressure is required to cause inertial cavitation *in vivo*. Nevertheless, an MI of 1.9 was set as an arbitrary regulatory threshold historically to standardize the application of diagnostic ultrasound. This pragmatic threshold has proven valuable in providing extensive evidence of the mechanical safety of diagnostic ultrasound when applied below this level. It is important to note that the reverse is not true: there is no evidence that mechanical or cavitation damage occurs immediately upon exceeding MI = 1.9. The true mechanical safety limit may be considerably higher, for example, damage by inertial cavitation with very short pulses has been observed at MI > 15 in the brain [13] and MI > 13 in a variety of tissues and tissue phantoms [14]. Cavitation probability, and indeed MI, rises with decreasing fundamental frequencies. A comprehensive review of MI and cavitation risk in the ultra-low-frequency domain, as low as 20 kHz, highlights the safe use of ultrasound levels below MI ≤ 1.9 [15].

MI was introduced theoretically for single diagnostic ultrasound pulses at higher frequencies (the duration of a single cycle, generally 0.1 to 0.5 µs) [16–19], while TUS often consists of considerably longer low-frequency (< 1 MHz) ultrasound pulses or trains of pulses. Nonetheless, there is a consensus that the MI ≤ 1.9 and $MI_{tc}$ ≤ 1.9 levels are acceptable for the longer TUS pulses and trains of pulses used for neuromodulation. Blackmore *et al.* [20] provided a summary of findings in small and large animals undergoing TUS, mostly below $MI_{tc}$ = 1.9, in which there is no convincing evidence of damage. More recently, Gaur *et al.* [21] provided further empirical data that at $MI_{tc}$ far exceeding 1.9, there is no evidence of damage. This is further supported by *in vivo* experimental measurements for one-second-long continuous wave (CW) sonications [22]. In a series of experiments, the threshold to induce cavitation through focused ultrasound was significantly above MI = 1.9, scaling with the center frequency at 5.3 MPa/MHz with a 0.6 MPa offset. Notably, this implies the margin between MI = 1.9 and the cavitation threshold also scales with the square root of frequency, with reduced margins at lower frequencies (e.g., 0.8 MPa or 95% margin at 200 kHz) and larger margins at higher frequencies (e.g., 1.9 MPa or 140% margin at 500 kHz). Overall, the lowest MI required to induce cavitation was observed for 246 kHz sonication at MI > 3.7. In summary, there is substantial evidence that an $MI_{tc}$ of 1.9 is below the cavitation threshold and non significant risk.

Note also that reaching a transcranial $MI_{tc}$ > 3.7 at 200 kHz, assuming a conservative derating of skull attenuation [23], requires a free-field pressure of at least 2.1 MPa. While this still provides some theoretical difference between the pressure for an $MI_{tc}$ of 1.9 and an $MI_{tc}$ of 3.7, care must be taken to control for standing waves as they might lead to local increases in pressure through constructive interference. Standing waves can be especially significant at low frequencies and with small-aperture transducers [24] or when focusing near bone [25]. Standing waves could amplify the in situ pressure and increase $MI_{tc}$, highlighting the need to consider standing waves to maintain the realized $MI_{tc}$ at or below 1.9 for a nonsignificant risk profile.

The MI is typically defined at the focus from measurements in free field. However, note that the peak-rarefactional pressure in transcranial applications may no longer be at the free-field peak but might occur elsewhere. For safety considerations, the consensus levels consider peak-rarefactional pressure in soft tissues, such as the scalp and brain, but exclude the skull. For transcranial applications with reasonably focused transducers, the pressure at the scalp could be expected to be significantly lower than at the free field focus. One possibility is that $MI_{tc}$ is defined not necessarily at the free field focus, but at the location of the *in situ* spatial peak pressure. Nonetheless, at all times, it remains the responsibility of the investigator, operator, or manufacturer to select the index most appropriate for the configuration, MI or $MI_{tc}$. In either situation, TUS applications with an appropriately calculated MI or $MI_{tc}$ below 1.9 can be considered non significant risk.

Beyond the risk of cavitation, we consider mechanical risks in relation to pressure, intensity, and strain. Indeed, in diagnostic ultrasound, safe domains of acoustic pressure are not only ensured by constraining MI, but also by constraining acoustic intensity. The FDA regulatory limit is that spatial-peak pulse-average intensity ($I_{sppa}$) be kept below 190 W/cm². However, in the frequency domain of TUS (< 1 MHz), MI will always be more constraining than $I_{sppa}$, rendering an additional threshold on acoustic intensity superfluous. Further, in diagnostic ultrasound, these pressure limits also ensure a non significant risk domain for tissue strain induced by particle displacement and radiation force. For example, this is supported by the safe application of strain elastography and shear wave elastography (SWE). Importantly, the mechanical stresses induced by TUS are of similar magnitude to those of diagnostic ultrasound, again emphasizing that a threshold on MI or $MI_{tc}$ is sufficient. However, when operating beyond our assumptions, such as with higher frequencies, it is possible that MI will not limit strain to NSR and further work is needed to define these strain thresholds [26]. In summary, we conclude that MI and $MI_{tc}$ are pragmatic and comprehensive metrics to constrain all types of mechanical effects for TUS in the absence of cavitation nuclei, such as ultrasound contrast agents.

> **TUS Safety Consideration 1.** Mechanical risks are nonsignificant if either the **MI or $MI_{tc}$** does not exceed 1.9, in the absence of cavitation nuclei, such as ultrasound contrast agents. One should determine which mechanical index corresponds best to the configuration: MI or $MI_{tc}$. For many TUS applications, this will be the $MI_{tc}$.

## 2.2. Thermal safety

The mechanical energy transported by focused ultrasound waves can be transferred to thermal energy through absorption, leading to tissue heating. The thermal energy deposition is proportional to the square of the pressure and is thus usually maximum at the focus in the absence of the skull. However, in the specific case of transcranial ultrasound stimulation, because of the high acoustic absorption of the skull, extra attention should be paid not only to the target but also to cortical locations close to the skull surface, even when stimulating targets deep in the brain. While skin, muscle, fat, and bone tissue are more resistant to thermal damage than brain tissue [27], we do not exclude skull and scalp tissue from our consensus. First, for transcranial applications, most heat can be expected to be deposited at the outer surface of the skull [28,29]. Further, the heat accumulated inside the skull will diffuse to the surrounding tissues, including the brain, and will radiate and transfer heat after the end of the ultrasound pulse. Indeed, it is important to consider sufficient diffusion time when estimating thermal risk. Below, we first review the existing guidelines for other biomedical devices and then discuss our consensus on NSR thermal levels for TUS.

### 2.2.1. Thermal safety of tissues in biomedical applications

The thermal rise in biological tissues, including the brain, induced by energy deposition, has been assessed for many other medical devices. Three guidelines are relevant for TUS:
- *Medical electrical equipment*: the international foundational standard for medical electrical equipment provides a framework of requirements for device safety and performance, including protection against excessive temperatures. The outer surface of external devices must not exceed 43°C when in continuous contact (IEC 60601-1 / Medical electrical equipment - Part 1: General requirements for basic safety and essential performance [30]).

- *Magnetic Resonance Imaging*: the maximum temperature of tissues should be 39°C when the MRI device is operated in normal mode (IEC 60601-2-33 / Particular requirements for the basic safety and essential performance of magnetic resonance equipment for medical diagnosis can be found here [31]).
- *Implantable devices*: an active implantable device should comply with one of two general thermal conditions or should be justified by evidence for a particular application:
    1. For implantable devices not intended to supply heat, no outer surface of the device shall be greater than 2°C above the normal surrounding body temperature of 37°C when implanted.
    2. No tissue receives a thermal dose exceeding the tissue-specific thresholds: 2 CEM43 for brain, 16 CEM43 for bone, and 21 CEM43 for skin tissue. (See the standard for a complete overview for all tissue types.)

(ISO 14708-1 and ISO 14708-3 / Details on Active implantable medical devices can be found here [32][33].

These guidelines for temperature rise have been developed in the context of global temperature rise (e.g., MRI without temperature monitoring) or chronic application with limited control to remove the source of thermal risk (e.g., active implanted devices). Thus, they inform a conservative worst-case level for potential focal heating by TUS.

It is commonly held that the assumption when the MRI regulation was made was that the brain temperature was equal to the core body temperature of 37°C. This is explicit in the implantable device standard.

Recent studies have inserted some controversy concerning the baseline temperature of the brain. Rzechorzek et al. [34] used MR spectroscopy (water to NAA peak) to show variations in brain temperature spatially (higher temperatures more centrally) and temporally (higher temperatures early in the day). They varied by age, sex, and luteal phase, with a mean brain temperature of 38.5°C. Sharma et al. [35] used MR spectroscopy (water to creatine peak) to measure a mean brain temperature of 37.2°C. Horiuchi et al. [36] used DWI to measure temperature in the lateral ventricles and found a mean temperature of 37.24°C in the morning and 37.11°C in the evening, showing the same daily trend as Rzechorzek, but with lower mean temperatures. It is unclear how to consider these temperatures, which were measured with a technique known to increase temperature (MRI). As long as these other standards hold, it is reasonable that TUS is in alignment with them, specifically, that the thermal risks are nonsignificant if the temperature rise is less than 2°C or the absolute temperature is 39°C, assuming a baseline temperature of 37°C.

> **TUS Safety Consideration 2.** Thermal risks are nonsignificant if the **temperature** rise is less than 2°C or the maximum absolute temperature is 39°C, assuming a baseline temperature of 37°C.

Temperature guidelines constrain the peak temperature, regardless of time and total exposure. To account for the confluence of both temperature magnitude and duration, the thermal dose is related to the integral of temperature over exposure time [37]. The dose of short exposure at a higher temperature might be equivalent to a longer exposure at a lower temperature. The thermal dose is expressed as a 'thermal isoeffective dose' in cumulative equivalent minutes (CEM) at 43 °C [37]. This metric is used to assess thermal safety in biomedical applications, including ultrasound exposure levels for diagnostic ultrasound [38], radiofrequency exposure levels for MRI [39], and thermal exposure levels for implantable devices [32]. The ISO 14708-3 standard specifies that the thermal dose formula is valid for temperatures between 39 °C and 57 °C [32].

Thermal sensitivity differs between tissue types, with higher thermal dose thresholds observed in tissues that are more commonly exposed to heat. For example, the threshold for thermal damage to skin tissue is observed at 210 CEM43 [38], i.e., an exposure time at a given temperature theoretically equivalent to an exposure time of 210 minutes at the reference temperature of 43 °C. The lowest value for thermal damage in the brain was found in dog tissue: 7.5 CEM43, [40]. This threshold was observed when the temperature of the brain was raised by infusing heated blood, thus establishing worst-case conditions by preventing any protective effect of perfusion and raising global rather than focal temperature.

It is recommended to adopt the lowest thermal dose threshold for the whole body by default, i.e., 2 CEM43, when tissue types are unknown. Tissue-specific thresholds can be adopted when tissue types can be mapped with confidence. This approach matches the AIUM consensus, recommending 10 CEM43 for short and 1 CEM43 for longer exposures, independent of tissue type [38]. Similarly, expert consensus on MR radiofrequency exposure has proposed to constrain CEM43 below 2 in healthy persons, i.e., with uncompromised thermoregulation, without the need for temperature-controlled conditions [39].

In summary, ITRUSST considers thermal risks nonsignificant if the thermal dose does not exceed 2 CEM43 in brain tissue, 16 CEM43 in bone tissue, and 21 CEM43 in skin tissue. We recommend these levels to align with the international standard ISO 14708-3 [32]. Note that the calculation of thermal dose encompasses the full duration during which tissue is exposed to temperatures exceeding 39 °C. This includes not only the sonication period, but also the period of cooling until 39 °C has been reached.

> **TUS Safety Consideration 3.** Thermal risks are nonsignificant if the **thermal dose** does not exceed 2 CEM43 in brain tissue, 16 CEM43 in bone tissue, and 21 CEM43 in skin tissue.

For the reader's interest, we give an example. One set of parameters that would result in 2 CEM43 would be 40°C for 128 min [37]. A thermal dose of 7.5 CEM43 would require, for the same temperature of 40°C, a duration of 480 min.

### 2.2.2. Thermal safety of tissues in diagnostic ultrasound

The Thermal Index (TI) was introduced in the AIUM/NEMA Output Display Standard [41] and incorporated by the FDA into guidelines for manufacturers to ensure clinical users are informed about potential thermal effects during diagnostic ultrasound. It was derived from an analysis of the thermal rise induced by acoustic energy deposition in [42].

Thermal indices have been defined for three different tissue types:

- *Soft tissue Thermal Index* (TIS) for general abdominal and peripheral vascular imaging. TIS assumes that only soft tissue is insonated.
- *Bone at Focus Thermal Index* (TIB) for obstetrics. TIB assumes bone is present at the depth where temporal intensity is greatest.
- *Cranial Thermal Index* (TIC) for adult transcranial scanning. TIC assumes the bone is very close to the front face of the diagnostic ultrasound probe. It further makes the conservative assumption of complete absorption of ultrasound energy at the bone surface.

Most TUS setups correspond to the TIC configuration, where the skull is close to the front face of the TUS transducer. The TIC is defined by IEC 62359:2010 [43]:

$$TIC = \frac{W_0/D_{eq}}{C_{TIC}}$$

where $C_{TIC}$ is a constant = 40mW.cm$^{-1}$ and $W_0$ is the transducer output power in mW. TIC was introduced for imaging applications where the planar face of the transducer is in direct contact with the scalp. Thus, $D_{eq}$ is usually set as the diameter (in cm) of the active aperture of the transducer. Appropriate determination of the TIC, as specified by IEC 62359:2010 [43], requires knowledge of the device's power output, which can and should be provided by the manufacturer. An example of a calculation of the TIC is given in Martin et al. [2]. For TUS application, most transducers are not in direct contact with the skin and the best practice is to define $D_{eq}$ as the aperture diameter of the acoustic beam on the outer surface of the skull in cm [2] (Fig. 1).

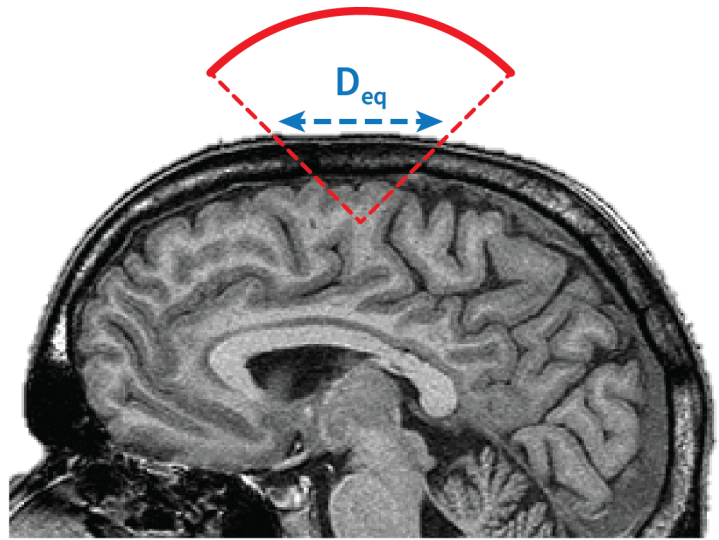

*Figure 1: Equivalent aperture of the beam for a typical TUS configuration.*

The international standard [43] provides a complete description of how to measure $W_0$. $W_0$ should be averaged over the whole neurostimulation exposure, but ITRUSST recommends that the averaging window should not exceed 30 seconds. The rationale for the 30-second limit is the characteristic diffusion time [44] of the skull bone.

Regarding the upper end of the TIC range given in Table 1:
- The FDA states that the manufacturer should explain the reason for any TI that exceeds a value of 6.0.
- BMUS does not recommend TIC > 3, TIB > 6, or TIS > 6. More precisely, for adult transcranial applications (imaging and standalone), BMUS recommends restricting exposure time to 60 min for 0.7 < TIC ≤ 1.0; 30 min for 1.0 < TIC ≤ 1.5; 15 min for 1.5 < TIC ≤ 2.0; 4 min for 2.0 < TIC ≤ 2.5; and 1 min for 2.5 < TIC ≤ 3.0 [45].
- For applications such as adult transcranial ultrasound, AIUM does not recommend TI > 6 and recommends restricting exposure time to 120 min for 1.5 < TI ≤ 2.0; 60 min for 2.0 < TI ≤ 2.5; 15 min for 2.5 < TI ≤ 3.0; 4 min for 3.0 < TI ≤ 4.0; 1 min for 4.0 < TI ≤ 5.0; and 15 s for 5.0 < TI ≤ 6.0 [46]. AIUM additionally states that as actual values of TIB are typically higher than displayed values for ARFI and pulsed Doppler, due to underestimation of intensity caused by spatial averaging of measurements of these pulses.[46] As thermal dose increases with exposure time, this effect may be countered by reducing maximum scanning times by 33% for ARFI and pulsed Doppler examinations when bone is near the transducer focus. Consequently, AIUM recommends restricting exposure time to 80 min for 1.5 < TI ≤ 2.0; 40 min for 2.0 < TI ≤ 2.5; 10 min for 2.5 < TI ≤ 3.0; 160 s for 3.0 < TI ≤ 4.0; 40 s for 4.0 < TI ≤ 5.0; and 10 s for 5.0 < TI ≤ 6.0 for ARFI and pulsed Doppler examinations when bone is near the transducer focus. Since these are the most conservative recommendations made by AIUM, we recommend these exposure duration limits are observed for TUS.

> **TUS Safety consideration 4.** Thermal risks are nonsignificant if the exposure time does not exceed 80 min for 1.5 < TI ≤ 2.0; 40 min for 2.0 < TI ≤ 2.5; 10 min for 2.5 < TI ≤ 3.0; 160 s for 3.0 < TI ≤ 4.0; 40s for 4.0 < TI ≤ 5.0; and 10 s for 5.0 < TI ≤ 6.0. One should determine which **thermal index** corresponds best to the configuration: TIS, TIB, or TIC. For many TUS applications, this will be the TIC.

The FDA considers the spatial-peak temporal-average intensity ($I_{spta}$). The $I_{spta}$ relates to the temporal-average exposure and thus thermal rise. However, as highlighted by the AIUM, the thermal dose and thermal indices are superior to estimate the thermal risk of ultrasound exposure [38,41]. ITRUSST concurs with the AIUM conclusions regarding the inferiority of the $I_{spta}$ compared to thermal dose and TI and, therefore, does not include $I_{spta}$ in the consensus metrics for thermal safety. The bioeffects that would be limited by $I_{spta}$ levels are already sufficiently constrained to non significant risk levels by the mechanical and thermal metrics given above.

## 3. Specific considerations for TUS
### 3.1. Insertion loss of the skull

Existing guidelines for calculating the mechanical index use 0.3 dB/cm/MHz as a derating factor of the peak-rarefaction pressure in the brain and do not consider the insertion loss induced by the skull bone [17].

The following methods could be used to better estimate the actual pressure achieved *in situ*, and thus derive the $MI_{tc}$ for transcranial applications. All methods must be either subject-specific or generalized to the population given a sufficiently large number of estimations.

- *Numerical models*: numerical models can be used to approximate acoustic pressure in the whole head, including reflection and propagation of the acoustic wave through the skull. This approach is best suited to consider the high variability in skull thickness between individuals and across stimulation sites for a given individual. It is recommended to collect subject-specific information on skull characteristics (e.g., using CT or MRI) to benefit the accuracy of the simulations. Alternatively, a simulation-based model could be derived from a group of skulls representing the population, and, as above, investigators should describe the model used to determine the estimation and estimate the uncertainty of their approach or justify why their measures reflect the inter-individual variability or why their model is conservative. Numerical models should have been validated with experimental measurements on a representative set of human skulls. Methods and best practices to perform numerical simulations are described in international guidelines and the literature [47–49]. While further work is needed on estimation of uncertainty in simulations, the existing literature on errors and uncertainties in transcranial simulations [50,51] and propagation of uncertainty [52] are useful starting points.
- *Analytical models*: the minimum impact of the skull can be estimated by considering the transmission loss at bone interfaces and a frequency-dependent loss in the bone. When using this approach, investigators should describe the model used to determine the estimation and quantify the uncertainty of their approach or justify why their measures reflect the inter-individual variability or why their model is conservative. One possible model is the three-layer model with absorption introduced recently [23]. Note that this analytical approach is not valid when ultrasound is focused on peripheral tissue, such as the skin, where the peak-rarefactional pressure is not expected to be transcranial. It remains the responsibility of the investigator, operator, or manufacturer to select the index and approach most appropriate for the configuration.
- *Experimental measurements*: a calibrated hydrophone can be used to map the acoustic beam reflected off and transmitted through a sample of human skulls or a sample of representative skull phantoms. When using this approach, investigators should describe the phantoms used to determine the estimation, and either quantify the uncertainty of their approach or justify how their measures and estimation of the transmission are conservative in the context of inter-individual variability in skull thickness, morphology, and composition. Measurements performed on one single skull cannot capture such variability. Guidance for acoustic measurement can be found in the ITRUSST reporting guidelines [2] and should be performed according to the most appropriate currently available standards [53,54].

*In situ* pressure estimates can be used to estimate the $MI_{tc}$ in the head. It may also be similarly informative to calculate an empirical estimate of the TIS using the estimated in situ focal intensity (noting that the spatial peak pressure or intensity may not always be at the intended target, depending on the transducer geometry, focal characteristics, positioning, skull morphology etc.). The standard definition of the TIS uses the $I_{spta}$ or acoustic power derated by 0.3 dB/cm/MHz, i.e. assuming the beam propagates through soft tissue only. However, a similar applicable quantity may be calculated using the estimated in situ $I_{spta}$.

### 3.2. Insertion loss of brain tissues

References for the attenuation of brain tissues cover a range [55–58]. We recommend the use of 0.5 dB/cm/MHz, with a linear dependence on frequency, as it is conservative compared to the empirical values provided in the literature.

### 3.3. Estimation of Temperature Rise

There are several methods that can be used to estimate the temperature rise in the skull and brain. Refined models can be derived with the inclusion of any of the following techniques:
- *Numerical models*: numerical models can be used to approximate the propagation of the acoustic wave through the skull and estimate the associated thermal rise. This approach is best suited to reflect the high variability in skull thickness between human subjects and across stimulation sites for a given subject. When using this approach, investigators should describe the model used to determine the estimation. Numerical models should have been validated with experimental measurements on a representative set of human skulls.
- *Analytical models*: the temperature should be appraised by estimating the thermal deposition in tissues. When using this approach, investigators **should describe the model used to determine the estimation.**
- *Experimental measurements*: temperature can be measured at the skull surface and at focus of the subject, such as with MR-thermometry. Alternatively, a model can be derived from representative head phantoms. Investigators should describe the phantom used to determine the estimation. Measurements should be performed on a large set of skulls, representative of the population of patients or volunteers to be stimulated.

For any of these approaches, the investigator must demonstrate that their approach is conservative in the context of the wide range of human head geometries.

## 4. Discussion

There are currently no regulatory guidelines for the non significant risk application of ultrasound neuromodulation in humans. Here, we establish expert consensus on parameters and levels that suggest a non significant risk operating regime without potential for serious risk. We are not making any statements about significant risk above the proposed levels of mechanical and thermal metrics. These levels are not intended to be limiting. Higher levels can be justified, for example, by appropriate monitoring of safety indices such as through passive cavitation detection or temperature measurement. In all cases, it is the responsibility of the investigator, operator, or manufacturer to provide adequate justification for the chosen approach. It is also their responsibility to comply with all the existing regulations concerning human subject research and medical device safety.

There are several excellent review articles evaluating the biophysical safety of TUS in human and animal studies [20,59–63][10][9]. To summarize, there are no reports of severe adverse events following TUS in humans for studies adhering to the thermal and mechanical metrics presented here. There are reports of mild/moderate adverse events, including headache, fatigue, mood deterioration, scalp heating, itchiness, neck pain, muscle twitches, anxiety, cognitive problems, and sleepiness. Only one study has looked at histological changes in tissue samples resected from human studies. In this study, evaluation of tissue from temporal lobe resection of epilepsy patients following sonications with MI of 2.14 and TIC of 2.88 found inconclusive or no detectable damage to the tissue in 7 patients and inconclusive findings in one patient [61]. In the animal literature, the reviews point to only a few studies using low-intensity ultrasound that described any potential damage after ultrasound. In Lee et al., histological examination of sheep demonstrated hemorrhage after ultrasound [64], but there was no discussion of any reactive tissue (inflammation associated with the hemorrhage), limiting the conclusion that the hemorrhage was from the ultrasound. In addition, the study did not include any control animals. Further, in a follow-up study, Gaur et al. [21] found equivalent hemorrhage in control animals and again found a lack of reactive tissue. Kim et al. [65] found that 1 of 30 rats had an incidence of hemorrhage after ultrasound, but again, there was neither a control set of animals nor a discussion about reactive tissue. In summary, there is no conclusive evidence of biophysical harm at the levels covered in the body of literature cited by these five review articles. Encouragingly, a large and growing body of literature consistently demonstrates that effective neuromodulation can be achieved within the conservative parameters specified here [66,67]. This suggests that there is a sufficiently large therapeutic window between the thresholds for effective neuromodulatory dose and significant biophysical risk [68]. While many of the studies covered operated within the levels specified by this consensus paper, several of the animal studies significantly exceeded these conservative levels (e.g., [21]) and did not show evidence of biophysical damage. This again underscores the non-significant risk of the levels specified here. An independent group of experts also proposed recommendations for TUS and came up with similar levels, with the same MI, TI and temperature limits based on the same IEC standards on MRI and implanted devices [69] previously highlighted in the first available version of our article [70]. Whether single cycle or multicycle ultrasound is used, the consensus is that any MI or $MI_{tc} \leq 1.9$ is considered NSR.

## Current limitations and future developments

The ITRUSST levels specified here are not safety limits; application beyond these levels and assumptions does not necessarily imply a significant risk. Further data is needed to establish the threshold for significant risk. We recognize some potential limitations of our approach to evaluating the relevance of existing regulatory guidelines to TUS application.

The ITRUSST consensus does not consider the presence of calcifications in the brain. These small calcium deposits can range in size from microscopic to macroscopic and are often present in the pineal gland and the choroid plexus, but also in the habenula, dura, and other brain structures [71]. Calcifications might absorb and aberrate acoustic waves. While microscopic calcifications are highly prevalent across the population, found in more than 70% of adults [71], macroscopic calcifications become especially prevalent with advancing age [72]. Importantly, calcifications are not a contraindication for diagnostic ultrasound [73], and no adverse events are associated with calcifications, including transcranial Doppler imaging, a modality with acoustic exposure in the same range as our consensus for TUS. For ablative high-intensity focused ultrasound of the brain, patients are screened for calcifications with CT, and the stimulating elements are adjusted to minimize acoustic energy at the calcifications.

We recognize the numerous observations supporting the nonsignificant risk of ultrasound delivered with a mechanical index below 1.9, including observations from TUS studies that consider MI for transcranial application *in situ*, at the target tissue in the brain [20]. However, constraining $MI_{tc}$ to the brain might not explicitly consider the potential for reflections and standing waves between the transducer surface and the skull. Indeed, MI is defined at the free field focus, agnostic of the skull, and $MI_{tc}$ is defined for all soft tissues. In conventional applications of TUS with sufficiently focal transducers, one could expect the pressures in the scalp to be significantly lower than at the free field focus. However, it is possible for standing waves to emerge in the scalp leading to a higher $MI_{tc}$. Investigations of low-frequency low-intensity ultrasound of the skin, as used in sonophoresis, found that cavitation in the skin itself was insignificant, but that cavitation of the coupling medium could lead to bioeffects on the skin surface, follicles, and pores [74]. It is, therefore, recommended to minimize and control cavitation in the coupling medium, for example, by degassing water-based media or using viscous media such as mineral or castor oil [3,15]. In summary, it is conceivable that with controlled coupling, the mechanical risk in the scalp is sufficiently minimized by adopting $MI_{tc}$ levels in the brain alone; however, this warrants further empirical quantification in configurations directly relevant to TUS, i.e. with transducers close to the scalp. Similarly, while cavitation has been investigated at low frequencies relevant to TUS (< 1 MHz) [22,74], most studies considered a higher frequency range (> 1 MHz) [13,14]. The field would benefit from further studies into the relationship between pressure, mechanical index, and cavitation threshold in the low-frequency domain of TUS [15].

The ITRUSST consensus for mechanical safety does not consider the skull tissue, while the consensus for thermal safety does not exclude the skull. These are deliberate choices to promote concise and pragmatic recommendations that are closely aligned with existing guidelines. Namely, skull tissue is not considered in the definition of MI, nor can pressure be straightforwardly measured inside the skull to obtain an empirical estimate of $MI_{tc}$ in this compartment. For similar reasons, the thermal considerations do not exclude the skull. First, bone tissue is not excluded from other medical device guidelines for thermal rise and thermal dose. Second, TIC explicitly considers the skull. Third, heating of the skull will diffuse into surrounding regions. By recommending consideration of skull temperature over a reasonable time window, we make this risk explicit.

Similar to a consensus for mechanical risk, ITRUSST adopts existing regulatory guidelines and standards for thermal risk. Implementing these guidelines appropriately remains the responsibility of the investigator, operator, or manufacturer. Several pragmatic choices can be made to aid this implementation. For example, several standards and guidelines define a peak absolute temperature for indefinite periods at 39 °C, while absolute baseline temperatures might not be known. A pragmatic implementation of this guideline would be to assume a baseline temperature of 37 °C in all tissues for healthy individuals and limit temperature rise to 2 °C. Further, while the ISO 14708-3 standard specifies thermal dose thresholds specific to tissue types, a conservative implementation would be to restrict thermal dose to the lowest defined limit, i.e., 2 CEM43, across all tissues, or at least across all intracranial tissues, avoiding the need to define and validate different tissue compartments in the intracranial cavity. Similarly, we recommend applying the limit for bone tissue (16 CEM43) to the full skull compartment and the limit for skin tissue (21 CEM43) to all tissue types in the scalp.

When multiple sonications are applied in series, the question arises as to whether subsequent exposures should be considered as a single or as separate thermal dosages. Two factors are of note here: first, whether the temperature has returned to normal (≤ 39 °C), and second, whether any transient thermal effects, including macromolecular changes in response to hyperthermia [75], have returned to baseline. By definition, the thermal dose calculation already considers the cooling period to normal temperatures (≤ 39 °C). However, the return to macromolecular baseline might outlast the cooling period. One conservative approach could be to accumulate the thermal dosage of all sonications that are part of one session or intervention. An alternative approach could be to consider the dose-dependent period until all transient hyperthermia-induced macromolecular changes have returned to baseline [76]. In this framework, one could conservatively consider subsequent sonications independent when they are separated by at least four times the period of the preceding thermal dose in equivalent minutes at 43 °C. For example, after an exposure equivalent to 2 CEM43, one could wait at least 8 minutes before starting a new sonication. This rationale is based on observations during the synthesis phase of mitosis, i.e., the phase most sensitive to heat-induced toxicity, suggesting a macromolecular recovery period equivalent to twice the preceding thermal dose at 43 °C for dosages up to 120 CEM43 [76].

The definition of thermal dose partially considers the effect of thermotolerance, where tissues are more resistant to thermal dose when applied at mild temperatures as mediated by heat-shock proteins [37]. Indeed, when calculating thermal dose, the exponent base R is set at 0.25 for temperatures < 43 °C and at 0.5 for higher temperatures [2]. Note that thermotolerance is not maintained beyond 43 °C, nor does it return upon cooling. In other words, once a temperature of 43 °C has been reached, the thermal dose should be calculated with the same base R of 0.5 for all subsequent time steps [37]. A conservative approach would be to limit peak temperatures to < 43 °C to simplify the calculation of thermal dose. This approach also ensures that any subsequent sonications can continue to benefit from the effect of thermotolerance.

We note that this manuscript has specifically addressed TUS in isolation. However, there is considerable interest and practice of combining TUS with MRI, each of which can similarly lead to a temperature rise. While not stated explicitly, the consensus here should include the MRI environment and include any potential temperature rise from the MRI. It is the ITRUSST consensus that it is non-significant risk if the combined temperature rise adheres to the thermal levels specified here.

The regulatory guidelines relevant to TUS, as reviewed here, do not differentiate between single, repeated, or chronic application. The same guidelines apply to all conditions. Indeed, when operating within these guidelines, no serious adverse events have been reported for repeated or long-term sonication (> 3 hours), including the application of transcranial Doppler in at-risk populations [77–79]. This is further corroborated by dedicated TUS safety studies, where no microstructural damage has been associated with TUS after repeated sonication in sheep (thousands of pulses over multiple days; [21]) and non-human primates (tens of sessions over multiple months; [80]), and without behavioral and physiological deficits in non-human primates (79 repeat sessions over multiple months; [81]). Accordingly, ITRUSST does not differentiate between single, repeated, or long-term sonication for biophysical safety and considers the risk for cumulative changes nonsignificant when sonication parameters do not exceed the consensus on NSR levels. However, even within the biophysically non significant risk range, one could consider the potential for physiological changes of neuromodulation that outlast the sonication period and might accumulate over time or repetition. Such effects might be a physiological safety consideration or, indeed, might be intended in the context of a clinical trial or treatment. Physiological safety considerations, including but not limited to, off-target neuromodulatory effects, cumulative neuromodulatory effects, adverse reactions, putative drug interactions, and inclusion and exclusion criteria, fall outside the scope of the current report on biophysical risk. These considerations will be addressed in an upcoming consensus report.

## 5. Summary

The International Consortium for Transcranial Ultrasound Stimulation Safety and Standards suggests levels of transcranial ultrasound stimulation to be nonsignificant risk for structural damage. They are not safety limits, but, at this point in time, there is insufficient data to provide a specific threshold for significant risk levels. These levels assume the application of TUS to persons who are not at risk for thermal or mechanical damage, without contraindications, and without ultrasound contrast agents. The current consensus aims to provide concise, comprehensive, and clear levels for biophysical safety. We anticipate that this document will be updated in line with emerging data on safety. In all cases, it remains the responsibility of an institutional review board or equivalent committee to make an ethical assessment that considers all risks and benefits, including potential intended and unintended physiological effects, contraindications, putative interactions, and physiological safety. It is likely that these aspects are co-dependent and will differ between health and disease. These considerations will be addressed in upcoming ITRUSST consensus meetings and reports.

**CRediT authorship contribution statement**

**Jean-François Aubry:** Conceptualization, Writing - Original Draft, Writing - Review & Editing, Visualization, Supervision. **David Attali:** Writing - Review & Editing. **Mark E. Schafer:** Writing - Review & Editing. **Elsa Fouragnan:** Writing - Review & Editing. **Charles F. Caskey:** Writing - Review & Editing. **Robert Chen:** Writing - Review & Editing. **Ghazaleh Darmani:** Writing - Review & Editing. **Ellen J. Bubrick:** Writing - Review & Editing. **Jérôme Sallet:** Writing - Review & Editing. **Christopher R. Butler:** Writing - Review & Editing. **Charlotte J. Stagg:** Writing - Review & Editing. **Miriam C. Klein-Flügge:** Writing - Review & Editing. **Seung-Schik Yoo:** Conceptualization, Writing - Review & Editing. **Christy K. Holland:** Writing – Review & Editing. **Brad E. Treeby:** Conceptualization, Writing - Review & Editing. Eleanor Martin: Writing – Review & Editing. **Lennart Verhagen:** Conceptualization, Writing - Original Draft, Writing - Review & Editing, Supervision. **Kim Butts Pauly:** Conceptualization, Writing - Original Draft, Writing - Review & Editing, Supervision.

**Funding acknowledgements**

JFA is supported by the Agence Nationale de la Recherche (ANR-20-CE19-0013 and ANR-10-EQPX-15), by the Technological Research Accelerator program of INSERM, and by the Focused Ultrasound Foundation (Center of Excellence program). JFA and DA are supported by the Bettencourt Schueller Foundation. EF is supported by a UKRI Medical Research Council Future Leaders Fellowship grant (MR/T023007/1), a BBSRC grant (BB/Y001494/1), and the Advanced Research + Invention Agency (ARIA). CFC is supported by the National Institute of Neurological Disorders and Stroke of the National Institutes of Health (UG3NS135551). EJB is supported by CURE Epilepsy and the Epilepsy Foundation New England. JS is supported by the Agence Nationale de la Recherche (ANR-2-CE37-0021). CRB is supported by the National Institute for Health and Care Research. CJS holds a Senior Research Fellowship, funded by the Wellcome Trust (224430/Z/21/Z). MKF is supported by Wellcome Trust grant 223263/Z/21/Z and UKRI grant EP/X021815/1. SSY was partially supported by Fund to Sustain Research Excellence from Brigham Research Institute. CKH is supported by a Lantheus™ research contract and holds the Hanna Endowed Chair of Cardiology at the University of Cincinnati. EM is supported by a UKRI Future Leaders Fellowship (MR/T019166/1), and by the EIC Pathfinder project CITRUS (Grant Agreement No. 101071008) funded by the EU Horizon Europe research and innovation program. LV is supported by a VIDI fellowship funded by the Dutch Research Council (NWO, 18919) and is a co-applicant on an EIC Pathfinder project funded by the European Innovation Council (EIC, 101071008) and on an ERC Advanced project funded by the European Research Council (ERC, MediCoDe). KBP is supported by grants from the National Institutes of Health (NIH R01 MH131684, NIH R01 NS112152, NIH R01 EB032743).
**Declaration of competing interest**

Jean-François Aubry reports a relationship with Insightec that includes: funding grants. Jean-François Aubry reports a relationship with Sonomind that includes: equity or stocks. Mark E. Schafer reports a relationship with BrainSonix Corporation that includes: consulting or advisory. Elsa Fouragnan reports a relationship with Attune Neurosciences that includes: board membership and consulting or advisory. Robert Chen reports a relationship with Attune Neurosciences that includes: consulting or advisory. Robert Chen is the editor-in-chief of Clinical Neurophysiology and an editorial board member of Brain Stimulation. Ellen J. Bubrick serves on the Board for the International Society for Therapeutic Ultrasound 2025-2027. Charlotte J. Stagg is a Deputy Editor at Brain Stimulation. . Christy K. Holland reports relationships with Lantheus™ and Boston Scientific Corp. which includes consulting and material transfer agreements. Bradley E. Treeby reports a relationship with NeuroHarmonics LTD that includes: board membership, employment, and equity or stocks. Eleanor Martin reports a relationship with Brainbox that includes: consulting or advisory. Eleanor Martin is a member of the Ultrasonics Technical Committee of the IEC. Lennart Verhagen reports a relationship with Brainbox that includes: non-financial support. Lennart Verhagen reports a relationship with Sonic Concepts that includes: non-financial support.  Lennart Verhagen reports a relationship with Image Guided Therapy that includes: non-financial support. Lennart Verhagen reports a relationship with Nudge that includes: consulting or advisory. Kim Butts Pauly reports a relationship with Attune Neurosciences that includes: consulting or advisory and equity or stocks. Kim Butts Pauly reports a relationship with Surf Therapeutics that includes: consulting or advisory. Kim Butts Pauly reports a relationship with MR Instruments that includes: non-financial support. If there are other authors, they declare that they have no known competing financial interests or personal relationships that could have appeared to influence the work reported in this paper.